\documentclass[amsmath,amssymb,superscriptaddress,nobalancelastpage,aps,prl,twocolumn]{revtex4-2}

\usepackage{graphicx}
\usepackage{dcolumn}
\usepackage{bm}
\usepackage{color}
\usepackage{pifont}
\usepackage{natbib}
\usepackage{hyperref}
\usepackage{capt-of}
\usepackage{amsmath} 
\usepackage{enumitem} 
\usepackage[normalem]{ulem}
\usepackage{braket}
\hypersetup{
colorlinks = true,
urlcolor   = blue,
linkcolor  = blue,
citecolor  = blue
}
\newcommand{\KCO}{\text{K}_2\text{Cr}_8\text{O}_{16}}

\begin{document}


\title{Topological Metal-Insulator Transition within the Ferromagnetic state}

\author{Ola~Kenji~Forslund}
   \email{ola.forslund@physik.uzh.ch}
\affiliation{Physik-Institut, Universität Zürich, Winterthurerstrasse 190, CH-8057 Zürich, Switzerland}
\affiliation{Department of Physics and Astronomy, Uppsala University, Box 516, SE-75120 Uppsala, Sweden}

\author{Chin~Shen~Ong}
 \email{chinshen.ong@physics.uu.se}
\affiliation{Department of Physics and Astronomy, Uppsala University, Box 516, SE-75120 Uppsala, Sweden}

\author{Moritz~M.~Hirschmann}
\affiliation{RIKEN Center for Emergent Matter Science (CEMS), Wako, Saitama 351-0198, Japan}

\author{Nicolas~Gauthier}
\affiliation{Institut National de la Recherche Scientifique – Énergie Matériaux Télécommunications, Varennes, QC J3X 1P7 Canada}

\author{Hiroshi~Uchiyama}
\affiliation{SPring-8/JASRI, Sayo, Hyogo 679-5198, Japan}

\author{Christian~Tzschaschel}
\affiliation{Physik-Institut, Universität Zürich, Winterthurerstrasse 190, CH-8057 Zürich, Switzerland}

\author{Daniel~G.~Mazzone}
\affiliation{Laboratory for Neutron Scattering and Imaging, Paul Scherrer Institut, CH-5232, Villigen, PSI, Switzerland}

\author{Romain~Sibille}
\affiliation{Laboratory for Neutron Scattering and Imaging, Paul Scherrer Institut, CH-5232, Villigen, PSI, Switzerland}

\author{Antonio~M.~dos~Santos}
\affiliation{Neutron Scattering Division, Oak Ridge National Laboratory, Oak Ridge, Tennessee 37831-6473, United States}

\author{Masafumi~Horio}
\affiliation{Physik-Institut, Universität Zürich, Winterthurerstrasse 190, CH-8057 Zürich, Switzerland}

\author{Elisabetta~Nocerino}
\affiliation{Laboratory for Neutron Scattering and Imaging, Paul Scherrer Institut, CH-5232 Villigen PSI, Switzerland}
\affiliation{Department of Materials and Environmental Chemistry, Stockholm University, SE-10691 Stockholm, Sweden}

\author{Nami~Matsubara}
\affiliation{Department of Applied Physics, KTH Royal Institute of Technology, SE-106 91 Stockholm, Sweden}

\author{Deepak~John~Mukkattukavil}
\affiliation{Department of Physics and Astronomy, Uppsala University, Box 516, SE-75120 Uppsala, Sweden}

\author{Konstantinos~Papadopoulos}
\affiliation{Department of Physics, Chalmers University of Technology, SE-41296 G\"oteborg}

\author{Kazuya~Kamazawa}
\affiliation{Neutron Science and Technology Center, Comprehensive Research Organization for Science and Society (CROSS), Tokai, Ibaraki 319-1106, Japan}

\author{Kazuhiko~Ikeuchi}
\affiliation{Neutron Science and Technology Center, Comprehensive Research Organization for Science and Society (CROSS), Tokai, Ibaraki 319-1106, Japan}

\author{Hidenori~Takagi}
\affiliation{Max Planck Institute for Solid State Research, Heisenbergstraße 1, 70569 Stuttgart, Germany}

\author{Masahiko~Isobe}
\affiliation{Max Planck Institute for Solid State Research, Heisenbergstraße 1, 70569 Stuttgart, Germany}

\author{Jun~Sugiyama}
\affiliation{Neutron Science and Technology Center, Comprehensive Research Organization for Science and Society (CROSS), Tokai, Ibaraki 319-1106, Japan}

\author{Johan~Chang}
\affiliation{Physik-Institut, Universität Zürich, Winterthurerstrasse 190, CH-8057 Zürich, Switzerland}

\author{Yasmine~Sassa}
\affiliation{Department of Applied Physics, KTH Royal Institute of Technology, SE-106 91 Stockholm, Sweden}

\author{Olle~Eriksson}
\email{olle.eriksson@physics.uu.se}
\affiliation{Department of Physics and Astronomy, Uppsala University, Box 516, SE-75120 Uppsala, Sweden}
\affiliation{Wallenberg Initiative Materials Science for Sustainability, Uppsala University,
75121 Uppsala, Sweden}

\author{Martin~M\aa nsson}
    \email{condmat@kth.se}
\affiliation{Department of Applied Physics, KTH Royal Institute of Technology, SE-106 91 Stockholm, Sweden}

\date{\today} 

\begin{abstract}
A major challenge in condensed matter physics is integrating topological phenomena with correlated electron physics to leverage both types of states for next-generation quantum devices. Metal-insulator transitions (MITs) are central to bridging these two domains while simultaneously serving as `on-off' switches for electronic states. 
Here, we demonstrate how the prototypical material of K$_2$Cr$_8$O$_{16}$ {\color{black}undergoes a ferromagnetic MIT accompanied by a change in band topology.} Through inelastic x-ray and neutron scattering experiments combined with first-principles theoretical calculations, {\color{black}we demonstrate that this transition is not driven by a Peierls mechanism, given the lack of phonon softening. Instead, we establish the transition as a topological MIT within the ferromagnetic phase (topological-FM-MIT)} with potential axionic properties, {\color{black} where electron correlations play a key role in stabilizing the insulating state}. This work pioneers the discovery of a topological-FM-MIT and represents a fundamentally new class of topological phase transitions, {\color{black}revealing a unique pathway through which magnetism, topology, and electronic correlations interact.}

\end{abstract}

\keywords{short keywords that describes your article}

\maketitle

A major milestone in condensed matter physics is bridging the gap between topological phenomena and correlated electron physics to harness both magnetic and topological states for next-generation quantum devices. Central to this effort are metal-insulator transitions (MITs), which serve as both conceptual and functional bridges between these two domains while simultaneously acting as ideal 'on-off' switches for manipulating electronic states -- an essential mechanism for device engineering.

While most known MITs occur in antiferromagnetic states with no net magnetization, ferromagnetic MITs -- which maintain net magnetization across the transition -- offer unique functionality for quantum information and spintronics. However, such transitions are exceptionally rare and poorly understood, particularly when involving topological character. Most well-studied topological phase transitions (TPTs) rely on the closing and reopening of a local gap at a Weyl point, characteristic of single-particle Weyl semimetals~\cite{Yan2017, Armitage2018, Hu2019}. In contrast, a more profound form of TPT involves a correlation-driven MIT, where the gap closes without reopening -- a phenomenon largely unexplored in topological systems. Such ferromagnetic MITs expand the frontier of topologically driven material manipulation, presenting unprecedented opportunities for innovation in systems where strong electron correlations intersect with topological phenomena~\cite{Kane2005, Bernevig2006, Fu2007, Hasan2010, Weng2015, Huang2015, Xu2015}. 

In a typical Weyl semimetal, Weyl points exhibit chiral symmetry, meaning points of opposite chirality do not interact~\cite{Wang2013}. Pairing interactions that break this symmetry, such as charge density waves (CDWs), can induce new phases with topological Goldstone modes known as axions. When $\bm{q}_{\rm CDW}$ nests Weyl points of opposite chirality near the Fermi level~\cite{Wang2013}, it can drive a TPT from a Weyl semimetal to a topological axion insulating phase, marking a MIT.

\begin{figure*}[ht]
  \begin{center}
    \includegraphics[keepaspectratio=true,width=\textwidth]
    {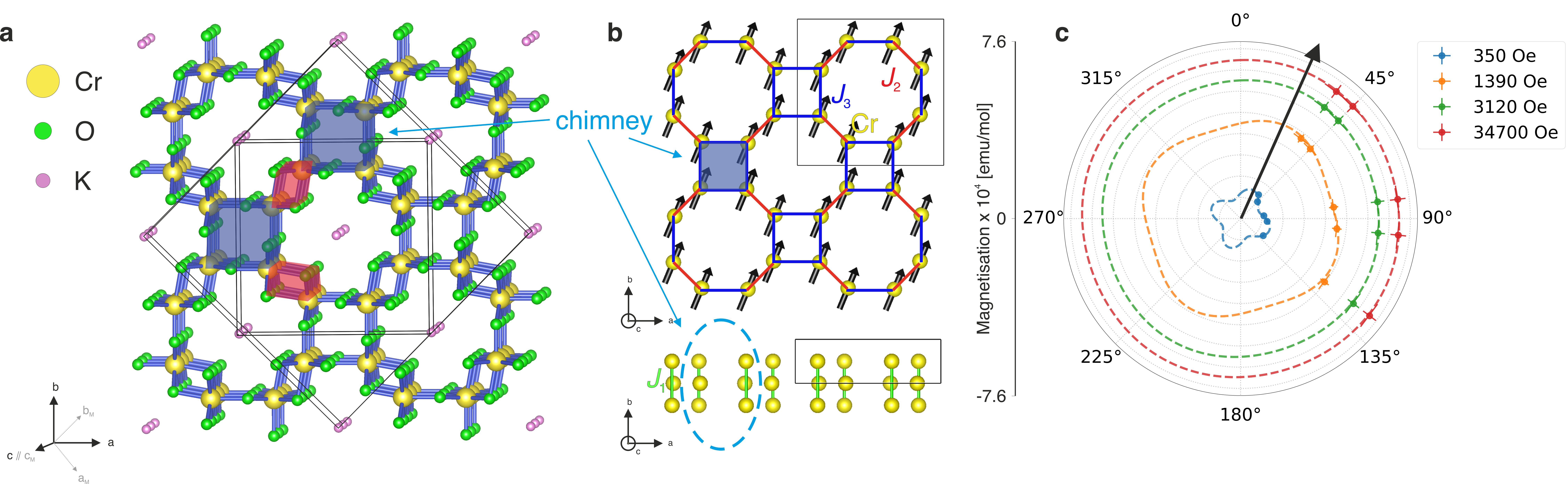}
  \end{center}
  \caption{{\bf The crystal structure of K$_2$Cr$_8$O$_{16}$.} (a) Tetragonal (smaller) and monoclinic (bigger) crystal structures of K$_2$Cr$_8$O$_{16}$, outlined by solid black lines. The shaded blue areas highlight the chimney building blocks, which comprise a square arrangement of four corner sharing CrO$_6$ octahedra that propagate along the c - axis. Here, $\mathbf{a}$ ($\mathbf{a}_\mathrm{M}$), $\mathbf{b}$ ($\mathbf{b}_\mathrm{M}$) and $\mathbf{c}$ ($\mathbf{c}_\mathrm{M}$) denotes lattice vectors of the tetragonal (monoclinic) phase. The shaded red areas highlight the double chain (inter-chimney) interactions made from Cr-Cr interacting through edge sharing CrO$_6$ octahedra. (b) Structure model showing only the magnetic Cr atoms. The bonding interactions, $J_2$ and $J_3$, are shown in the top panel, while $J_1$ is shown in the bottom panel. The chimneys are made up of $J_3$ and $J_1$ and represent intra-chimney interactions while $J_2$ is the inter-chimney interaction. {\color{black}The magnetic structure determined from ND and angle dependent magnetisation is included in (b) as black arrows. (c) Angle-dependent magnetisation at $T=120$~K, plotted in units of emu/mol for several applied magnetic fields (see legend).  Dashed lines represent sinusoidal fits to the data for each applied magnetic field. The arrow marks the easy axis orientation, approximately $22.5^\circ$ relative to the principal axis ($0^\circ$).
}
  }
  \label{fig:Crystal}
\end{figure*}

Traditional models, such as the Mott-Hubbard model, have successfully described MITs in various materials~\cite{Hubbard1964, Hubbard1964_2}, including high-temperature cuprate superconductors~\cite{Lee2006} or Verwey transitions in magnetites~\cite{Shchennikov2009}. However, understanding the temperature-dependent ferromagnetic-metal-to-ferromagnetic-insulator transition (FM-MIT) has proven particularly challenging. K$_2$Cr$_8$O$_{16}$ is renowned for exhibiting this FM-MIT. It has a Curie temperature $T_{\rm C}=167$~K with a MIT ($T_{\rm MIT}=95$~K) that retains the FM order~\cite{Hasegawa2009}. Since K$_2$Cr$_8$O$_{16}$ is considered a quasi-1D compound (Fig.~\ref{fig:Crystal}), the MIT has been explained by a 1D CDW~\cite{Toriyama2011}, also known as a Peierls transition, accompanied by a structural distortion from a metallic tetragonal phase to an insulating monoclinic phase (Fig.~\ref{fig:Crystal}) with $\bm{q}_{\rm CDW} = (1/2, 1/2, 0)$~\cite{Toriyama2011, Nakao2012}. However, {\color{black}a} recent study~\cite{Zhao2020} suggesting the presence of Weyl fermions raises new questions regarding the role of topological effects in this transition. {\color{black}Related compounds such as RbCr$_4$O$_{8}$, have also shown to host Weyl points~\cite{Xia2019}, demonstrating the robustness of these topological features in this family of compounds.}

\begin{figure*}[ht]
  \begin{center}
    \includegraphics[keepaspectratio=true,width=\textwidth]
    {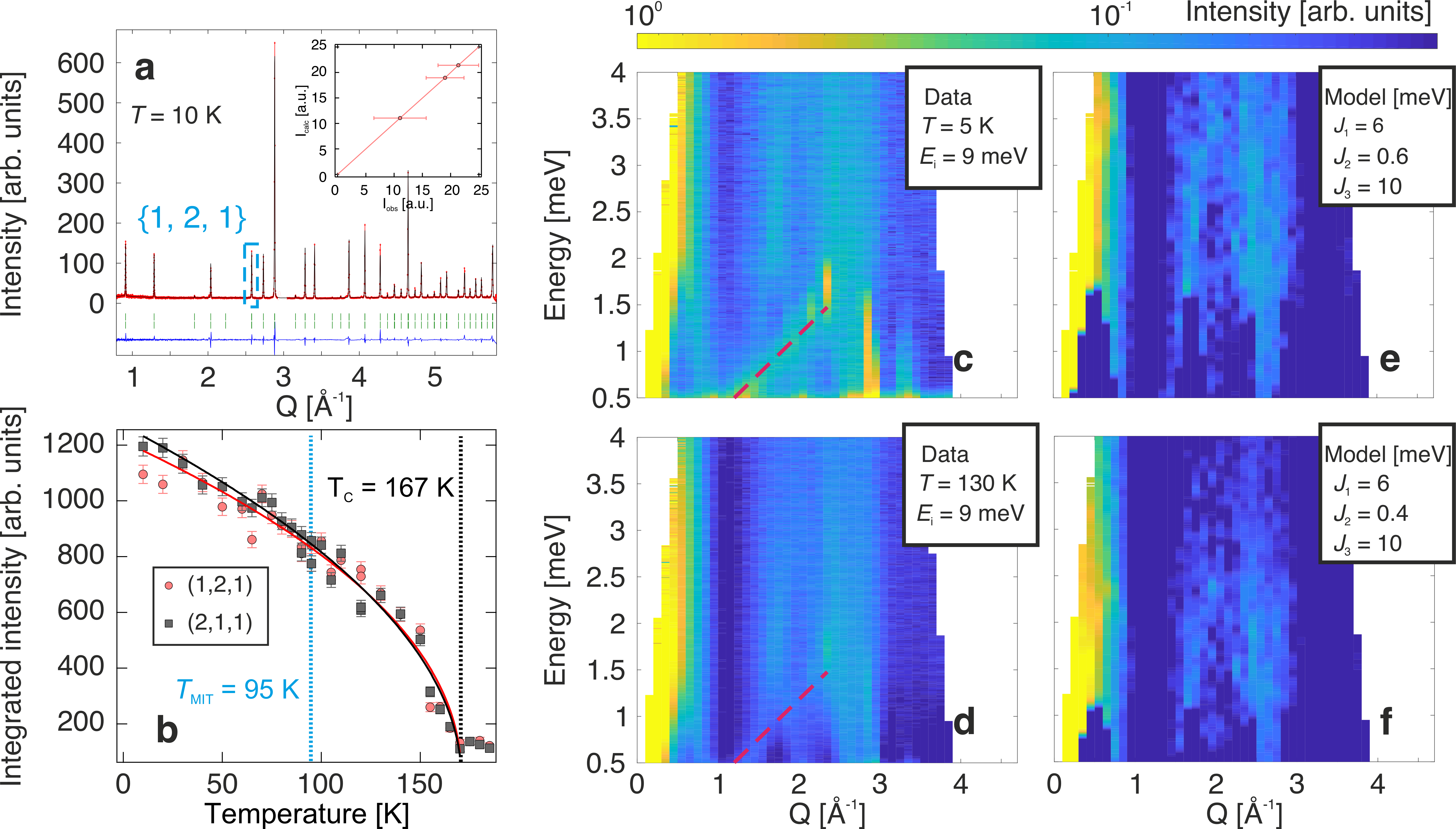}
  \end{center}
  \caption{{\bf Neutron scattering data of K$_2$Cr$_8$O$_{16}$.} (a) Neutron powder diffraction pattern collected at $T=10$~K, including the structural and magnetic refinements. The green ticks denote the allowed reflections and the blue line represents the difference between the model and the data. The main magnetic peak \{1,2,1\} is highlighted in blue (see SM Fig.~S3 for high temperature patterns). The inset shows a comparison between measured ($I_{\rm obs}$) and calculated ($I_{\rm calc}$) intensities from single-crystal magnetic refinement on datasets obtained from subtracting a selection of peaks at 10~K from the peaks at 200~K. (b) The (2,1,1) and (1,2,1) neutron diffraction peak intensities as a function of temperature. The solid line corresponds to the best fit using $I=I_0(1-T/T_{\rm C})^\beta$. (c, d) Collected powder inelastic neutron scattering spectra with the incident energy $E_i=9$~meV at 5~K and 130~K, respectively. The red dashed lines highlight the {\color{black}spurious-unknown} dispersion discussed in SM. (e,f) The best fit results obtained using the Heisenberg Hamiltonian for 5~K and 130~K. Fitting and analysis procedures are described in SM.}
  \label{fig:NS}
\end{figure*}

Using a combination of neutron diffraction (ND) and first-principles calculations, we demonstrate that the Weyl fermions of opposite chiralities are in fact nested by $\bm q_{\rm CDW}$. Furthermore, contrary to previous prediction of K$_2$Cr$_8$O$_{16}$ undergoing a Peierls transition for its FM-MIT~\cite{Toriyama2011}, our inelastic x-ray scattering (IXS) clearly shows the absence any phonon condensation, an experimental result that aligns with our first-principles calculations. This demonstrates that the FM-MIT is not driven by a Peierls transition. Instead, the FM-MIT of K$_2$Cr$_8$O$_{16}$ is a TPT, that is possibly axionic in nature. This work introduces a new class of TPT: a topological MIT within the FM order (topological-FM-MIT), paving the way for future quantum devices that leverage both magnetic and topological states. 
\\
\\
\noindent\textbf{\label{sec:result}Magnetic response across the transition.} As the positions of the Weyl points are highly dependent on the FM ordering~\cite{Zhao2020}, we investigated the magnetic response across the phase transitions using powder and single-crystal neutron scattering. The collected ND data and the corresponding Rietveld refinement at $T=10$~K are shown in Fig.~\ref{fig:NS}(a). Upon cooling below $T_{\rm C}=167$~K, a strong enhancement of the \{1,2,1\} reflections, consistent with additional scattering resulting from long range FM  ordering with $\bm{q_{\rm FM}} =(0,0,0)$ where $\bm{q_{\rm FM}}$ is the magnetic propagation vector. 
In order to identify the magnetic structure, the group-subgroup relationships for the given magnetic propagation were analyzed (Sec.~S2 of Supplementary Materials (SM)), and the solution of $C2'/m'$ (\#12.62) was obtained (Fig.~\ref{fig:Crystal}(b)). {\color{black} The FM spin polarization reduces the symmetry from the paramagnetic space group $I4/m$ (\#87) to the magnetic subgroup $C2'/m'$ (\#12.62), which is monoclinic.} The moment direction within the $ab$-plane could not be determined due to the powder average, an ambiguity that could not be resolved even with single-crystal ND due to the formation of magnetic domains. The magnetic \{1,2,1\} peaks could in principle be inequivalent below the temperature of the MIT ($T_{\rm MIT}$), due to the monoclinic distortion. However, temperature-dependent measurements of the peaks (1,2,1) and (2,1,1) do not show a difference (Fig.~\ref{fig:NS}(b)). {\color{black}To resolve the magnetic order within the $ab$-plane, angle dependent magnetisation measurements as a function of magnetic field at $T = 120$~K is presented in Fig.~\ref{fig:Crystal}(c). Fitting the data to a sinusoidal function yields an order direction tilted $\sim 22.5^\circ$ relative to the principal axis. Together, these results show that the magnetic order remains unchanged across the MIT, while the moment orientation at 120~K lies slightly off-axis within the $ab$-plane.}

To further study the spin correlations, inelastic neutron scattering (INS) spectra were measured on powder at $T = 5$ and 130~K (Fig.~\ref{fig:NS}(c,d)). Both spectra contained the same essential features, confirming the absence of any significant change in the exchange interactions across the MIT. Figure~\ref{fig:Crystal}(b) defines the most relevant exchange parameters for the compound: $J_1$ and $J_2$ (between nearest neighbor separated by $\sim2.9$~\AA) are both intra-chain interactions within the double chain of edge-sharing octahedra (Fig.~\ref{fig:Crystal}(a), shaded red), whereas $J_3$ (between third-nearest neighbors separated by $\sim3.4$~\AA) is the inter-chain interaction between corner-sharing octahedra (Fig.~\ref{fig:Crystal}(a), shaded blue). In this paper, we use the following conventions for the classical Heisenberg Hamiltonian, $\mathcal{H}$, defined as $\mathcal{H} = - \frac{1}{2} \sum_{ij} J_{ij} (\bm{e_i} \cdot \bm{e_j})$, where $\bm{e_i}$ is the unit vector in the direction of the $i$-th site magnetization and $J_{ij}$ is the isotropic exchange interaction parameter multiplied with the square of the magnetic moment of the $i$-th site. With this convention, we note that positive exchange values ($J$) favor FM alignment. \cite{Szilva2023}

Using linear spin wave theory~\cite{Kubler2021} and the magnetic order derived from ND, the exchange parameters, $J_1 =6.0$, $J_2 = 0.6$ and $J_3 = 10$~meV are obtained by fitting the experimental INS spectra to the Heisenberg Hamiltonian (Fig.~\ref{fig:NS}(e),(f) and Sec.~S2 of SM), consistent with FM spin alignment. The data also confirm that the MIT has little or no effect on the exchange couplings. To verify the experimentally obtained values, we used density functional theory (DFT) and the magnetic force theorem~\cite{Liechtenstein1984,Liechtenstein1987} to calculate $J$ from first principles. The theoretically calculated exchange parameters are also smaller between the nearest neighbors, $J_1$ and $J_2$, than they are for the third exchange parameter, $J_3$ (Fig.~S10(a) SM), in agreement with our experimental data and previous calculations~\cite{Toriyama2011}. These results confirm that the magnetic building blocks of the quasi-1D structure are the groups of four corner sharing chains between next-nearest neighbors (identified as the ``chimneys"~\cite{Toriyama2011}; Fig.~\ref{fig:Crystal}(a),(b) in shaded blue), instead of a one-dimensional linear chain (Fig.~\ref{fig:Crystal}(a),(b)). This questions the previous conclusions that $\KCO$ experiences Peierls instability at MIT, a notion predicated on $\KCO$ being 1D~\cite{Toriyama2011} and is further weakened by the absence of experimental evidence. We will return to this point later in this paper.

\begin{figure*}[ht]
  \begin{center}
    \includegraphics[keepaspectratio=true,width=1.00\textwidth]
    {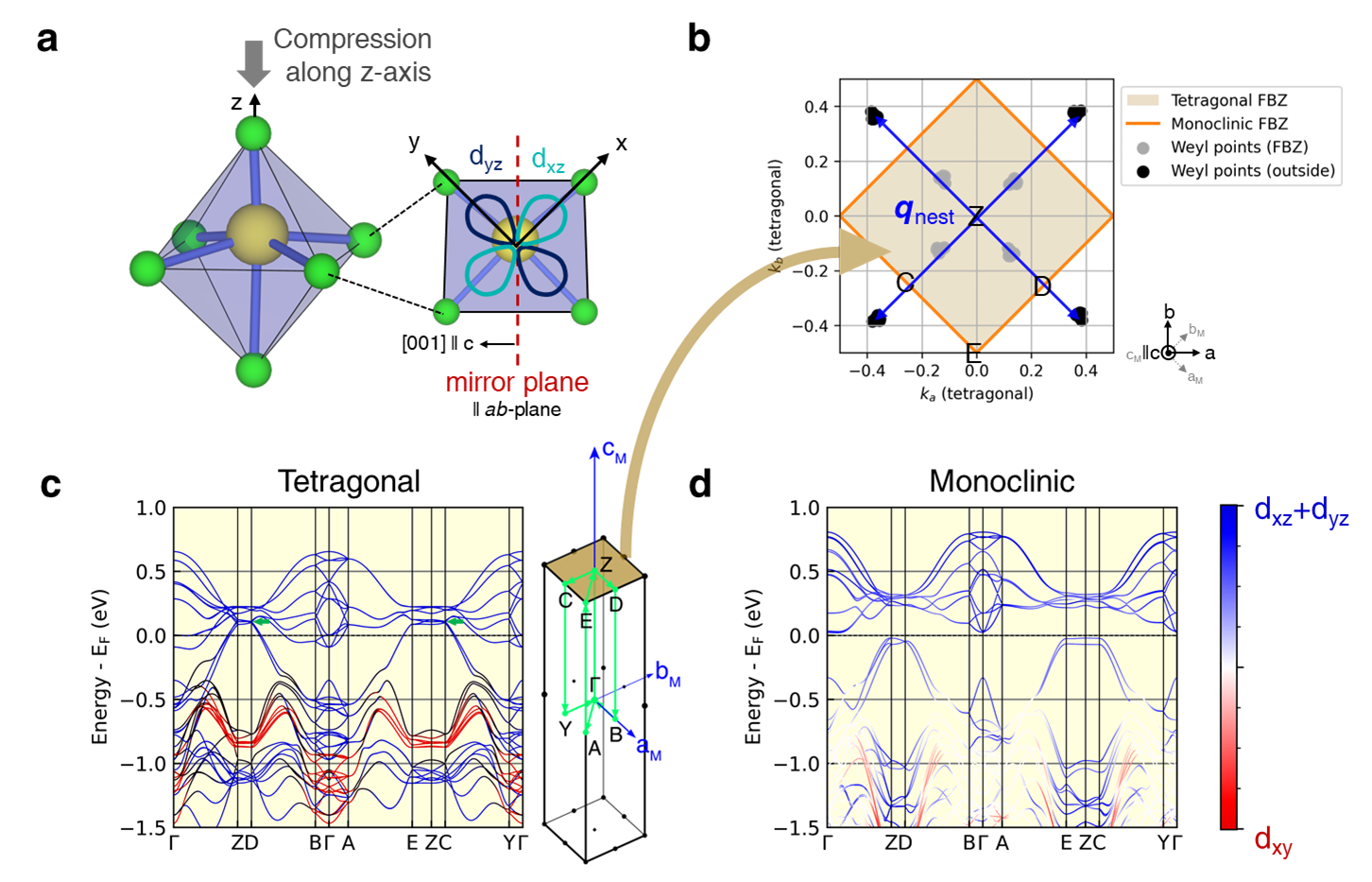}
  \end{center}
  \caption{{\bf First-principles theoretical calculations.} (a) Shows one of the distorted Cr-O octahedra that is compressed along the $z$-direction and the mirror symmetry is parallel to the $ab$-plane. $x$, $y$ and $z$ refer to the \textit{local} axes of the oc\-ta\-he\-dron. (b) {\color{black}Dots} denote the pairs of Weyl points in the first Brillouin zone of the tetragonal phase calculated in the monoclinic lattice supercell, defined by the $\mathbf{a}_\mathrm{M}$, $\mathbf{b}_\mathrm{M}$ and $\mathbf{c}_\mathrm{M}$ lattice vectors, located on the {\color{black}nodal plane of $k_c = \pi/c$}. The vector, $\bm q_{\rm nest}$, shows how the Weyl points are nested. (c,d) show the DFT band structures of the tetragonal {\color{black}($U=0.0$~eV)} and  monoclinic {\color{black}($U=4.0$~eV)} phases, respectively, with respect to the Fermi level ($E_F$). The color bar represents their projections on the $t_{2g}$-like Wannier orbitals. In (c), the small green arrow points to the approximate location of one of the Weyl points. Its inset shows the first BZ of the monoclinic lattice and the $\mathbf{k}$-path along which the band structures (in c,d) are plotted, with the shaded surface being the $k_c = \pi/c$ plane as shown in (b). } 
  \label{fig:theory}
\end{figure*}

{\color{black}
Finally, we found a clear $t^2/U$ dependence in the calculated exchange interactions, where $t$ is the strength of electron hopping between sites and $U$ is the Hubbard repulsion of electrons at the same site. This quadratic scaling indicates that the average interactions in K$_2$Cr$_8$O$_{16}$ arise from a superexchange mechanism~\cite{koch2012, Kvashnin2016}, rather than double exchange~\cite{Sakamaki2009, Sakamaki2010, Toriyama2011, Nakao2012, Nishimoto2012} -- for which $J_{\rm DE}\propto t$ -- or direct exchange type~\cite{Khomskii1997}. While superexchange is often associated with antiferromagnetism, in this case the dominant couplings are ferromagnetic, consistent with the Goodenough–Kanamori–Anderson rules. The microscopic origin is outlined in Sec.~S3C of the SM.}
\\
\\
\noindent\textbf{\label{sec:result}Topological phase transition.} Using structural~\cite{Tamada1996, Toriyama2011} and magnetic information derived from experiments, we analyzed the electronic properties by performing DFT calculations and employing a group theory analysis. We start our discussion with the paramagnetic tetragonal {\color{black}metallic phase}, with space group of $I4/m$ (\#87). In terms of atomic arrangement and structural periodicity, all Cr are symmetrically equivalent. The mirror symmetry relates their $d_{xz}$ and $d_{yz}$ orbitals to each other (Fig.~\ref{fig:theory}a), $i.e.$, in the paramagnetic phase all bands have an equal amount of $d_{xz}$ and $d_{yz}$ characters (Fig.~\ref{fig:theory}(c,d)) (as pointed out in Ref.~\onlinecite{Sakamaki2009}). In the present work, the spatial orientation of the $d$-orbitals are defined with respect to the local $z$-axis of the Cr-O octahedron (as defined in Fig.~\ref{fig:theory}(a)), which due to the crystal symmetry, always lies parallel to the $ab$-plane. The symmetry-equivalent $d_{xz}$ and $d_{yz}$ orbitals are higher in energy than the $d_{xy}$ orbital due to compression along the $z$-axis of the distorted Cr-O octahedra (Figs.~\ref{fig:theory}(a,c,d)). Since Cr in $\KCO$ has a nominal charge of $+3.75$, only ($6-3.75=$)~$2.25$ of its $t_{2g}$-like (i.e., $d_{xy}$, $d_{yz}$ and $d_{xz}$) orbitals are occupied. The fractional occupation of these states guarantees that the tetragonal phase is metallic or at least filling-enforced semimetallic (Fig.~\ref{fig:theory}(c)). 


In order to evaluate the topological nature of the bands, we used the DFT wavefunctions to construct a downfolded tight-binding model with the Cr $d_{xy}$, $d_{yz}$ and $d_{xz}$ Wannier functions as basis states. {\color{black}By calculating the Berry curvature of the magnetic metallic phase (\#12.62), we found pairs of Weyl points located close to $k_c=\pm\pi/c$ nodal plane 0.1~eV above the Fermi level (see Fig.~\ref{fig:theory}(b,c) and see Table~S3 SM for their precise locations)),} confirming the topological nature of the electronic structure. For each pair, Weyl points of opposite chiralities are related by inversion symmetry, consistent with the magnetic space group \#12.62. Furthermore, since all irreducible co-representations of this group are one-dimensional, the Weyl points are accidental crossings near the $k_c=\pm\pi/c$ plane, which explains why the Weyl points are not located at high-symmetry $\mathbf{k}$-points in the BZ (see Table~S3 SM for the locations). {\color{black}In addition, we find that the positions of the nodal points are robust. Regardless of the direction of the magnetic moment, we found that the nodal points always form a cross-like feature about the nodal plane, such that nested nodal pairs are always present in the $(0.5,0.5,0.0)$ and $(-0.5,-0.5,0.0)$ directions of the tetragonal Brillouin zone (SM  Fig.~S8).}

These Weyl points exhibit remarkable characteristics. {\color{black}In particular, the nesting vectors connecting Weyl points of opposite chirality, $\bm q_{\rm nest} = (0.75, 0.75, 0.0)$ and $(-0.75, -0.75, 0.0)$ (Fig.~\ref{fig:theory}(b)), closely match the lattice distortion wavevector observed in single-crystal X-ray diffraction experiments~\cite{Toriyama2011, Nakao2012}. This suggests that the nesting of Weyl points may be coupled to the lattice distortion, thereby breaking chiral symmetry and allowing the emergence of axion-like excitations~\cite{Gooth2019, Kaushik21, Ray2022}. A definitive confirmation of this scenario remains an important direction for future investigation.}

\begin{figure*}[ht]
  \begin{center}
    \includegraphics[keepaspectratio=true,width=\textwidth] 
    {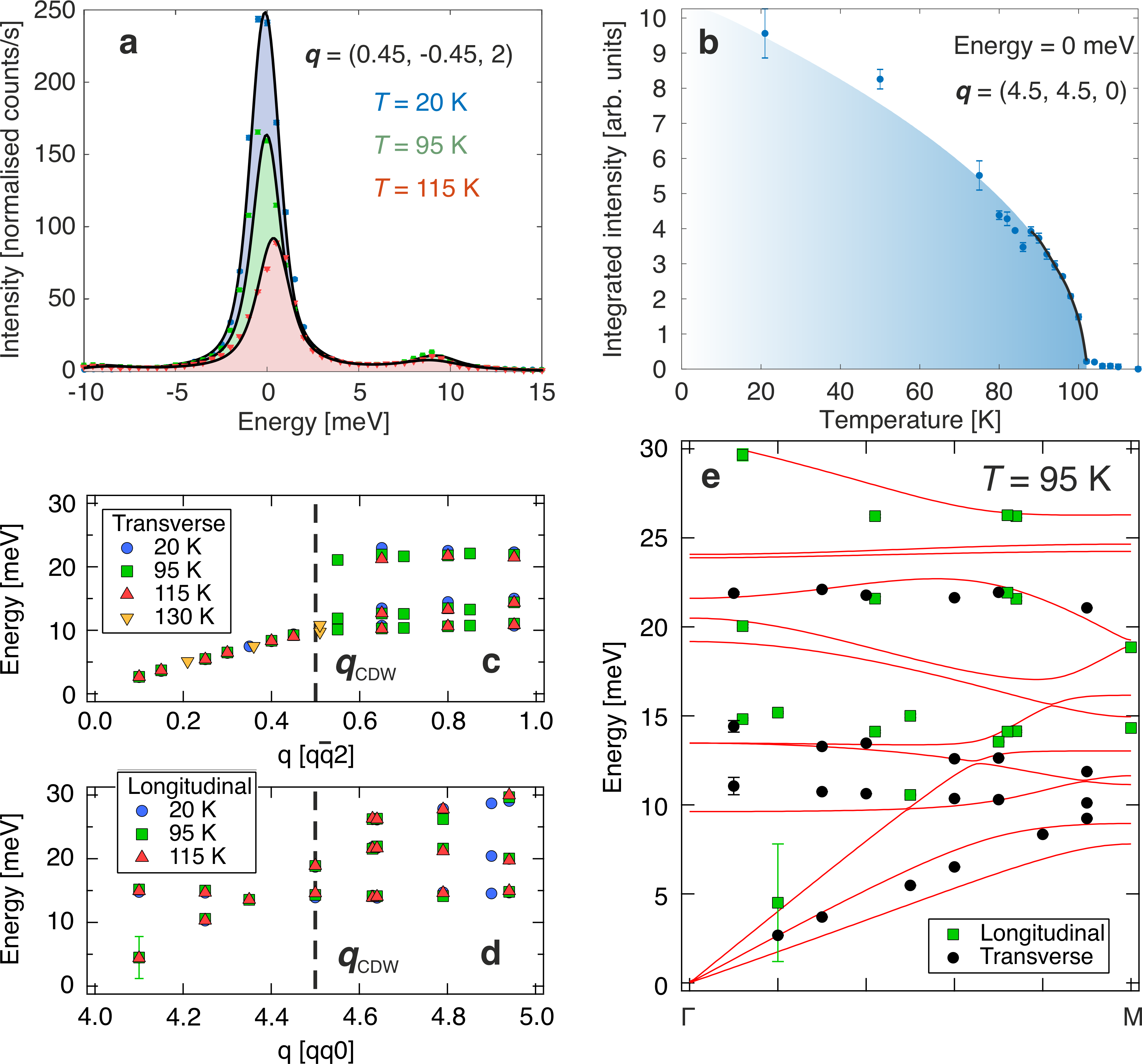}
  \end{center}
\caption{{\bf Inelastic x-ray scattering data of K$_2$Cr$_8$O$_{16}$.} (a) The inelastic x-ray scattering spectra collected in transverse geometry at 20~K (blue), 95~K (green) and 115~K (red). The scans were performed from $E=-10$~to 30 meV at $\bm q=h\bar{h}2$ for $h$=0.45. The solid black line represents the overall fit (Sec.~1 in SM). (b) The integrated elastic scattering intensity collected along $\bm q=h\bar{h}0$ between $h = 4.45$ and 4.55, $i.e.$ across $\bm q_{\rm CDW}$. Four scans were made and the errorbar represents the standard deviation. The solid line is a fit using the mean field expression: $I(T)=I_0(1-T/T_{C})^\beta$. (c, d) Phonon dispersion along $q\bar{q}l$, collected at $T=$~20, 95, 115~K in transverse and longitudinal geometries, respectively. One scan performed at 130~K is included as well. (e) Experimentally measured (symbols) and calculated (solid red lines) phonon dispersion from $\Gamma$ to M in folded configuration. Green and black symbols represent the longitudinal and transverse phonon modes, respectively.
  }
  \label{fig:IXS_DFT}
\end{figure*}

{\color{black}
Secondly, the Weyl points lie near the nodal planes at the Brillouin-zone boundary, $k_c = \pm\pi/c$, where every Bloch state is symmetry-enforced to be doubly degenerate. This degeneracy arises from the magnetic screw rotation $\mathcal{T}C_2^{[001]}$, which combines a 180° rotation about the $[001]$ axis with time-reversal symmetry. The operation squares to a lattice translation along $\mathbf{c}$ with a phase factor, $(\mathcal{T}C_2^{[001]})^2 = e^{ik_cc}$, so at $k_c = \pm\pi/c$, it imposes $(\mathcal{T}C_2^{[001]})^2 = -1$, enforcing degeneracy. This protection is a direct consequence of the base-centering in the conventional unit cell, which introduces a fractional translation in the symmetry operation. As a result, the $k_c = \pm\pi/c$ planes host symmetry-enforced nodal degeneracies that constrain the electronic band structure and promote the formation of Weyl points.

Having discussed the {\color{black}metallic} phase, we now proceed to analyze the {\color{black}insulating} phase, which belongs to the structural space group $P2_1/c$ (\#14). In this phase, the degeneracy between the $d_{xz}$ and $d_{yz}$ orbitals is lifted due to monoclinic distortions. Moreover, all four Cr atoms within each Cr$_4$O$_6$ chimney become inequivalent, further splitting the $d$-orbital levels into twelve non-degenerate states. As a result, the $d$ electrons can redistribute unevenly among the Cr sites. A nominal filling of 2.25 electrons per Cr corresponds to 9 out of 12 available $t_{2g}$-like states being occupied. {\color{black} This is consistent with the calculated (2.1~$\mu_B$) and experimentally measured (2.2~$\mu_B$) Cr magnetic moments. Moreover, the total electron count and the total spin moment per primitive cell are both integers, a condition necessary for the opening of a band gap.}

Since the magnetic moment directions are confined to the $ab$-plane, we deduced the highest symmetry magnetic subgroup to be $P2_1'/c'$ (\#14.79). This subgroup is different from metallic phase magnetic group $C2'/m'$ (\#12.62) in one distinct way: the monoclinic subgroup of the insulating phase is non-symmorphic, and its primitive unit cell coincides with the conventional one. Consequently, the screw rotation’s fractional translation does not arise from lattice centering, and the symmetry-enforced degeneracy on the $k_c = \pm \pi/c$ planes originates directly from the non-symmorphic symmetry, not from band folding. 

This reduction of translation symmetry is accompanied by a change in topological character. In the metallic phase, accidental crossings of the Cr $t_{2g}$-derived bands, allowed by the magnetic space group (\#12.62), produce Weyl points near the $k_c = \pm \pi/c$ planes. In the insulating phase, the Weyl points are removed due to the combination of monoclinic crystal distortion and orbital splitting. We further note that the nodal planes are topologically uncharged: they are symmetry-enforced by $\mathcal{T}C_2^{[001]}$, and inversion symmetry renders their Chern number trivial. By calculating the parity eigenvalues at all time-reversal invariant momentum points~\cite{Khalaf2018}, we confirm that the insulating monoclinic phase is a trivial insulator (i.e., $\mathbb{Z}_4$ index = 0). In summary, the disappearance of Weyl points is a direct consequence of translation symmetry reduction due to the tetragonal-to-monoclinic structural distortion, establishing the MIT as a topological phase transition within the FM state.} 
\\
\\
\noindent\textbf{\label{sec:electron}Phonon soft mode analysis.} We next focus on the driving mechanism behind the FM-MIT. The Peierls instability proposition is based on the experimental observation of superstructure peaks at $\bm q_{\rm CDW}=(0.5, 0.5, 0)$ below $T_{\rm MIT}$~\cite{Toriyama2011, Nakao2012}. We confirm the existence of these by plotting the temperature dependence of the integrated elastic peak intensity at $\bm q=h \bar h 0$ for $h = 4.50(5)$ in Fig.~\ref{fig:IXS_DFT}(b). In a Peierls transition \cite{Peierls1996}, the phonon energy of a strictly 1D material softens via the electron-phonon coupling due to divergence in the electronic susceptibility. This divergence causes the phonons energy to renormalise and collapse onto the elastic line (also known as a Kohn anomaly or phonon condensation), forming new Bragg peaks. 



In order to determine the role of phonons in a FM-MIT, we present IXS spectra collected at three different temperatures (115, 95 and 20~K): above, around and below the $T_{\rm MIT}$, for a wavevector close to $\bm q_{\rm CDW}$ ($\bm q=h\bar{h}2$ where $h=4.45$) in Fig.~\ref{fig:IXS_DFT}(a). Apart from the elastic peak, the energy scans contain inelastic Stokes and the smaller anti-Stokes pairs, attributed to a phonon mode. The observed phonon mode around $\bm q_{\rm CDW}$ reveal no temperature dependence ($\sim$10~meV for all temperatures close to $\bm q_{\rm CDW}$, Sec.~S1). This demonstrates the absence of phonon softening at $\bm q_{\rm CDW}$ during the MIT, firmly establishing that the phase transition is not a Peierls transition. 

The absence of phonon softening is more clearly shown in Fig.~\ref{fig:IXS_DFT}(c,d), where the phonon dispersion of $\bm q$-vectors across $\bm q_{\rm CDW}$ is shown over a wide range. 
The transverse acoustic phonon branch dispersed linearly with $q$ from the $\Gamma$-point for all temperatures. A similar linear dispersion was observed for the longitudinal acoustic phonon, albeit at a larger gradient with respect to $\bm q$. Nonetheless, irrespective of phonon polarization, we were unable to detect any phonon condensation at $\bm q_{\rm CDW}$. 
This is further corroborated by our first principles phonon calculations. The calculated phonon dispersion for the tetragonal structure from $\Gamma$ to M is superposed with experimental data points obtained from IXS with good agreement (Fig.~\ref{fig:IXS_DFT}(e)). These results are indirectly supported by a high-pressure experiment, which shows an antiferromagnetic ordering stabilized under pressure~\cite{Forslund2019}, instead of a FM-like ordering as theoretically predicted~\cite{Nishimoto2012} for a Peierls insulating FM compound. 

We confirm that there is no drastic phonon softening from a higher-energy optical phonons through our Density-Functional Perturbation Theory (DFPT) calculation. Instead, phonon modes in the {\color{black}insulating monoclinic} phase exhibit some minor softening for the low-energy Cr phonons relative to the {\color{black}metallic} phase (Fig.~S11(e)), leading to higher vibrational entropy. At low temperatures, the {\color{black}insulating} phase is favored over the {\color{black}metallic} phase as the former will have a lower Gibbs free energy.
The fully calculated phonon spectra containing all phonon branches are included in Fig.~S11(a,b).

Finally, a hard x-ray photoelectron spectroscopy (HAXPES) measurement~\cite{Bhobe2015} have showed that in the metallic phase, the valence of Cr is not static at the nominal valence value of 3.75. Instead, a dynamic noninteger Cr$^{4+}$-Cr$^{3+}$ valence fluctuations was observed. 
In fact, similar valence fluctuations have indeed mediated phase transitions such as structural instabilities in YbPb, where lattice-valence fluctuations coupling allowed a structural instability~\cite{Tsutsui2020}. One may speculate that there exists a valence-lattice coupling in K$_2$Cr$_8$O$_{16}$ as well, from which the low temperature structure is derived. 
%
\\
\\
\noindent\textbf{\label{sec:conclusions}Conclusion}\\
The driving mechanism behind the temperature dependent FM-MIT has remained elusive ever since its discovery 15 years ago~\cite{Hasegawa2009}. The quasi-1D nature of the $\KCO$ inspired theories invoking Peierls instability to explain the FM-MIT. Peierls instability hinges critically on phonon condensation \cite{peierls1955}, a phenomena disproved in this study. Instead, by combining state-of-the-art experimental work with theory, we have clarified that FM-MIT is in fact a topological phase transitions. Across the MIT, the distorted octahedral complex breaks the degeneracy between $d_{xz}$ and the $d_{yz}$ orbitals, thereby opening a gap. Consequently, $\KCO$ exhibits a unique transition where a singular structural distortion responsible for the MIT, simultaneously also destroys its topological phase, thereby realising a topological MIT within the FM state (topological-FM-MIT). To our knowledge, this work represents the first demonstration of such a topological-FM-MIT, bridging the realms of strong electron correlations and topological physics. This discovery lays the groundwork for future research on quantum devices that leverage both magnetic and topological states.
\bibliography{Refs}
\noindent\textbf{\label{sec:conclusions}Methods}
\\
\textbf{\label{sec:methods}Sample synthesis.} For this work, two single crystals ($0.5\times0.5\times1$~mm$^3$ and $0.1\times0.1\times0.2$~mm$^3$) with a $T_{\rm MIT}\simeq100~K$ were successfully prepared. The difficulty in determining the underlying physics of K$_2$Cr$_8$O$_{16}$ is partly related to the fact that the obtained single crystals are very small (associated with the necessary high-pressure synthesis protocol~\cite{Hasegawa2009}). We have referenced the value of $T_{\rm MIT}$ to 95~K in this work, based on the reported value of Ref.~\cite{Hasegawa2009}. {\color{black}In particular, new cell designs for a high-pressure Walker-type multianvil system were developed at the Max Planck Institute for Solid State Research, enabling growth of larger, higher-quality crystals. The new 18/11 and 25/15 multianvil cell assemblies provide more uniform pressure and temperature distributions, resulting in increased sample volumes (up to several mm in size) while maintaining stable synthesis conditions up to $\sim12$~GPa and $1500~^\circ$C. With these improved cells the CrO$_2$ impurity phase is reduced to below 2~\% in powder, and the obtained crystals are significantly larger (up to $0.5\times0.5\times1$~mm$^3$). The synthesis conditions otherwise follow those reported previously~\cite{Hasegawa2009}.} \\
\\
%
\textbf{\label{sec:methods}Inelastic X-ray scattering measurements.} The crystals were mounted onto separate copper sample holders and aligned to measure the transverse and longitudinal phonons using IXS. The IXS data were collected at the high resolution beamline BL35XU at the SPring-8 synchrotron source in Japan using $E_i=21.74$~keV. The inelastic peaks were fitted using a damped harmonic oscillator convoluted with the experimental resolution ($\sim1.7$~meV). The peaks were fitted using the software Dave~\cite{Dave}.
\\
\\
\textbf{\label{sec:methods}Magnetisation measurements.} 
{\color{black}
Angle-dependent magnetisation measurements were performed on a single crystal of $K_2$Cr$_8$O$_{16}$ (mass $m = 0.00148$~g) using a Quantum Design MPMS3 magnetometer in DC mode. Measurements were taken at selected angles between $0^\circ$ and $135^\circ$ (37°, 45°, 83°, 96°, and 127°) relative to the crystallographic $a$- and $b$-axes at $T = 120$~K under zero-field-cooled conditions. The initial crystal orientation was determined by Laue diffraction, after which controlled relative rotations were applied. The magnet was demagnetised between each angular measurement to minimize remanent fields. Magnetic fields up to 3.5~T were applied, and the measured magnetic moments were converted to molar magnetisation. Assuming fourfold rotational symmetry, the angular dependence of the magnetisation was fitted with the sinusoidal function; $M(\theta)=A\sin(4\theta+\phi)+M_0$.}
\\
\\
\textbf{\label{sec:methods}Neutron scattering measurements.} Neutron powder diffraction (NPD) was measured at the Spallation Neutron Source (SNS) in Oak Ridge National Lab (ORNL) using the POWGEN instrument~\cite{Huq2019}, set at a central wavelength of 0.8~\AA and a bandwidth of 1~\AA. These measurements were made in high resolution mode. About 500~mg of sample was loaded under helium gas for temperature equilibration onto a vanadium. Single-crystal neutron diffraction (ND) was performed at the Paul Scherrer Institut (PSI) using the ZEBRA instrument with wavelength~$= 1.178$~\AA~on a single crystal (0.5 x 0.5 x 1 mm$^3$). A 4 circle geometry mode with a closed cycled refrigerator was used in order to measure selected peaks from 10~K up to room temperature. The inelastic neutron scattering (INS) was measured at Japan Proton Accelerator Research Complex (J-PARC) using the 4SEASON instrument on a powder sample of $\sim2$~g with a resolution of $5\%$ of $E_i$~\cite{4season1,4season2,Ryoichi2011}. Given the uniqueness of the instrument, several $E_i$s (7, 9, 12, 18, 27, 47, 97 and 300~meV) were measured simultaneously at 10, 130 and 200~K and the relevant data is presented in this work. All measured neutron scattering intensities were normalised to the spallation source proton current. The diffraction pattern was analyzed using the Fullprof software package~\cite{Fullprof} while the inelastic spectra were analysed using SpinW~\cite{Toth2015}.\\
\\
\textbf{\label{sec:methods}Density functional theory calculations.} The DFT calculations were performed using the \texttt{Quantum Espresso}~\cite{Giannozzi2009} package. {\color{black} All DFT calculations for bandstructures, $J_{ij}$, phonon, Berry curvature and Weyl point analyses were performed in the ferromagnetic spin-polarized state. DFT calculations were carried out with $U=4.0$~eV for the insualting monoclinic phase to capture electron-correlation effects, whereas $U=0.0$~eV was used for the metallic phase, where metallic screening is expected to strongly reduce the effective on-site interactions.} The phonon eigenvectors and eigenvalues were calculated using the finite displacement and supercell approach with \texttt{Phonopy}~\cite{Phonopy}. Wannier functions are constructed from the DFT eigenvectors and eigenvalues using \texttt{Wannier90}~\cite{Pizzi2020}. The Cr $t_{2g}$ states were used as initial projections in the local coordinates of the octahedral crystal field. Using these Wannier functions, we use the magnetic force theorem~\cite{Liechtenstein1984,Liechtenstein1987} to map the tight-binding Wannier Hamiltonian onto a classical Heisenberg model using \texttt{TB2J}~\cite{He2021}. The chiralities of the Weyl points were calculated using \texttt{WannierTools}~\cite{wu2017} with the same Wannier basis sets. For more technical details, see Sec.~S3 of SM.\\
\\
\noindent\textbf{\label{sec:conclusions}Author contributions}\\
The project and the experiments were conceived by O.K.F, with valuable input from N.G, D.G.M, and M.M. The experiments were conducted by O.K.F, N.G, H.U, {\color{black}C.T.}, D.G.M, M.H., N.M, E.N, D.J.M, K.P, K.K, K.I, R.S, A.M.S, J.S, J.C, Y.S and M.M. The theoretical calculations were conceived by C.S.O and conducted by C.S.O, M.M.H. and O.E. The samples were synthesised by M.I and H.T. O.K.F and C.S.O prepared the manuscript and all co-authors contributed to the final draft.
\\~\\
\noindent\textbf{\label{sec:conclusions}Competing interest}\\
The authors declare no competing interests. \\
\\
\noindent\textbf{\label{sec:conclusions}Data availability}\\
All data needed to evaluate the conclusions in the paper are present in the paper. Additional data requests should be addressed to the corresponding authors.\\
\\
\noindent\textbf{\label{sec:conclusions}Acknowledgments}\\
The IXS measurements were performed at the SPring-8 with the approval of the JASRI (Proposal No. 2019A1525 and 2020A1653). One of the neutron scattering experiments was performed at the MLF, J-PARC, Japan, under a user program (No. 2019A0287). A portion of this research used resources at the Spallation Neutron Source (under the IPTS 21093), a DOE Office of Science User Facility operated by the Oak Ridge National Laboratory. This research was supported by the Swedish Research Council (VR) via a Neutron Project Grant (Dnr. 2021-06157), a geenral project grant (Dnr. 2022-03936), as well as the Swedish Fondation for Strategic Research (SSF) within the Swedish national graduate school in neutron scattering (SwedNess), and the Carl Tryggers Foundation for Scientific Research (CTS-22:2374). O.K.F is supported by the Swedish Research Council (VR) through Grant 2022-06217, the Foundation Blanceflor fellow scholarships for 2023 and 2024, and the Ruth and Nils-Erik Stenbäck Foundation. M.~M.~H.~is funded by the Deutsche Forschungsgemeinschaft (DFG, German Research Foundation) - project number 518238332. E.N. acknowledges financial support from the SSF-Swedness grant SNP21-0004 and the Foundation Blanceflor 2024 fellow scholarship. Y.S. is funded by the Knut and Alice Wallenberg FOndation through the Wallenberg Academic Fellow program (KAW 2021.0150). J.S. acknowledges support from Japan Society for the Promotion Science (JSPS) KAKENHI Grants No. JP18H01863 and No. JP20K21149. O.E. and C.S.O. acknowledge support from the Swedish Research Council (VR), the Knut and Alice Wallenberg Foundation (KAW), ERC (synergy grant FASTCORR, project 854843), Wallenberg Initiative Materials Sciencefor Sustainability (WISE) funded by the Knut and Alice Wallenberg Foundation, eSSENCE, and STandUPP. We also would like to acknowledge the Swedish National Infrastructure for Computing (SNIC) for computational support. All crystal structure figures were made with VESTA software~\cite{Vesta2008}.


\newpage

\newpage

\newpage

\end{document}